\documentstyle[11pt]{article}

\begin{document}
\newcommand{\BE}{\begin{equation}}
\newcommand{\EE}{\end{equation}}
\newcommand{\BA}{\begin{eqnarray}}
\newcommand{\EA}{\end{eqnarray}}
\newcommand{\pri}{\>'}
\setcounter{page}{1}
\begin{flushright}
{\small DOE/ER/05096-52}\\
{\small Revised version}
\end{flushright}
\vspace{24mm}
\begin{center}
{\LARGE\bf
Autonomous Renormalization of the
\vspace{1mm}\\
One-Loop Effective Potential
\vspace{5mm}\\
and a 2 TeV Higgs in $SU(2) \times U(1)$}
\vspace{28mm}\\
{\Large R. Iba\~nez-Meier \hspace*{2mm} and \hspace*{2mm} P. M. Stevenson}
\vspace{12mm}\\
{\large\it
T. W. Bonner Nuclear Laboratory, Physics Department,
\\
Rice University, Houston, TX 77251 (USA)}
\vspace{38mm}\\
\end{center}
{\bf Abstract:}
Branchina {\it et al} and Consoli have recently shown that the one-loop
effective potential (1LEP) of massless $\lambda \phi^4$ theory can be
renormalized in two distinct ways.  One of these is the conventional
renormalization of Coleman and Weinberg. The other requires an infinite
wavefunction renormalization, and is very similar to the ``autonomous"
renormalization of the Gaussian effective potential.  We apply the
``autonomous'' renormalization to the 1LEP of the $SU(2) \times U(1)$
electroweak theory with zero bare Higgs mass.  The predicted
physical Higgs mass is 1.9 TeV.
\vspace{4 mm}\\
%{\small PACS:
%03.65.Db, 11.10.Ef, 11.15.T
%}
\newpage
Only recently has it been realized that the familiar 1-loop effective
potential (1LEP) can be renormalized in two distinct ways \cite{bran,consoli}.
The renormalization-group (RG) equation satisfied by the 1LEP of massless
$(\lambda \phi^4)_4$ theory is compatible {\it either} with the positive,
perturbative $\beta$ function and an anomalous dimension
$\gamma = {\rm O}(\lambda^2)$ (as in Coleman and Weinberg \cite{cole})
{\it or} with a negative $\beta$ function and $\gamma = \beta/(2 \lambda)$.
The latter possibility \cite{bran,consoli}, which directly parallels the
so-called ``autonomous'' renormalization of the Gaussian effective potential
(GEP) \cite{con,aut,cast,kerman}, corresponds to an asymptotically free theory
with an infinite wavefunction renormalization (re-scaling of the classical
field) such that $\lambda \phi^2$ is an RG-invariant combination.  As stressed
by Refs. \cite{bran,consoli}, this approach suggests that the Higgs mass is of
order 2 TeV.  They also argue that its width will be relatively narrow
\cite{bran,consoli}.

     Some immediate comments are in order:  The conventional, perturbative
picture does not allow a Higgs mass greater than about 1 TeV \cite{lee}.
This is not a ``bound'' on the Higgs mass: Rather, it is statement about
where the perturbative description breaks down.  A heavy Higgs, in the
conventional view, has a large width due to large
Higgs-to-longitudinal-$W,Z$ couplings -- which are the manifestation of
large self-couplings in the scalar sector.  In the picture envisaged here
\cite{bran,consoli} the scalar sector, while it has a non-trivial
effective potential, has its self-interactions effectively ``squelched''
to zero by an infinite wavefunction renormalization.  Thus, there would be
{\it no} large Higgs-to-longitudinal-$W,Z$ couplings, and the Higgs would
be relatively narrow.  We discuss this a little further at the end.

In this letter, we describe the ``autonomous'' renormalization of the 1LEP
in a simple fashion, and show that it generalizes to any renormalizable
theory of scalars, fermions, and gauge bosons that is ``classically
scale-invariant'' (CSI) (that is, all terms in the Lagrangian have
dimension 4).  The CSI property means that all physical masses arise from
spontaneous symmetry breaking (SSB), and are proportional to the vacuum value
of the scalar field.  We consider in particular the $SU(2) \times U(1)$
electroweak theory, making the hypothesis that the bare Higgs mass is zero.
This hypothesis was once very popular \cite{cole,gild,ellis}, especially
in connection with the gauge-hierarchy problem, but went out of favour when
its apparent consequence, a 10 GeV Higgs, became ruled out by experiment.
However, with the ``autonomous'' renormalization, as we shall show, the
physical Higgs mass is predicted to be either 10 GeV, or 1.9 TeV.

\vspace*{3mm}
{\bf 1.} We begin by discussing the simple case of massless $\lambda \phi^4$
theory:
\BE
{\cal L}_{Bare} = \frac{1}{2} \partial_{\mu} \phi \partial^{\mu} \phi
- \lambda_B \phi^4.
\EE
The unrenormalized 1LEP is simply
\BE
V_{1l}= \lambda_B \varphi_c^4 + I_1(\sqrt{12 \lambda_B \varphi_c^2}),
\EE
where $I_1(\Omega)$ is the quartically divergent (Euclidean) integral
\BE
I_1(\Omega) \equiv \frac{1}{2} \int \frac{d^4 p}{(2 \pi)^4}
\ln (p^2 + \Omega^2),
\EE
which corresponds (up to an $\Omega$-independent, infinite constant) to the
vacuum energy of a free field of mass $\Omega$. A convenient formula for
$I_1(\Omega)$ is
\BE
\label{eye1}
I_1(\Omega) = I_1(0) +\frac{1}{2} \Omega^2 I_0(0)
- \frac{1}{8} \Omega^4 I_{-1}(\mu) + f(\Omega^2),
\EE
where $I_0$ and $I_{-1}$ are, respectively, quadratically and logarithmically
divergent integrals:
\BE
I_{0}(\Omega) \equiv  \int \frac{d^4 p}{(2 \pi)^4} (p^2 + \Omega^2)^{-1},
\EE
\BE
I_{-1}(\Omega) \equiv 2 \int \frac{d^4 p}{(2 \pi)^4} (p^2 + \Omega^2)^{-2},
\EE
and the finite function $f(\Omega^2)$ is given by
\BE
f(\Omega^2) \equiv \frac{\Omega^4}{64 \pi^2}
\left [ \ln \left( \frac {\Omega^2}{\mu^2}\right ) - \frac{3}{2} \right ].
\EE
This formula, (\ref{eye1}), is basically a Maclaurin expansion of $I_1$ in
powers of $\Omega^2$, except for the complication that ``$I_{-1}(0)$'' is
infrared divergent.  It is easily derived from the formulas in Ref.
\cite{gepii}, Sect. IIIB, which are themselves are obtained by
Taylor-expanding the $I_n$ integrals about $\Omega^2 = \mu^2$, where $\mu$
is some finite mass scale.  Note that the $I_n$ integrals, which we have
defined here as 4-dimensional Euclidean integrals, also have a simple form
as 3-dimensional integrals over spatial momenta \cite{gepii}.

In dimensional regularization $I_1(0)=I_0(0)=0$, so effectively
\BE
\label{eye1s}
I_1(\Omega) = - \frac{1}{8} \Omega^4 I_{-1}(\mu) + f(\Omega^2),
\EE
with $I_{-1}$ containing the divergent $1/\epsilon$ pole term.
Thus, one can write:
\BE
\label{ol2}
V_{1l} = \lambda_B \varphi_c^4 [1 - 18 \lambda_B I_{-1}(\mu)] +
f(12\lambda_B \varphi_c^2).
\EE
Conventionally, this would be renormalized by writing $\lambda_B = \lambda_R
(1 + 18 \lambda_R I_{-1}(\mu) + ...)$ and dropping ``higher-loop'' terms
\cite{cole}.  However, one can also obtain a finite, non-trivial result
with an ``autonomous'' renormalization in which the bare coupling constant
is infinitesimal and the classical field is infinitely re-scaled
(Cf. \cite{aut}):
\BE
\label{autren}
\lambda_B = \frac{\hat \lambda}{Z_\phi}, \qquad
\varphi_c= Z_\phi^{\frac{1}{2}}\Phi_c,
\EE
where $Z_\phi$ is proportional to $I_{-1}(\mu)$.
Inserting into (\ref{ol2}), the potential becomes:
\BE
V_{1l}= Z_{\phi} \Phi_c^4
\left [\hat \lambda - 18 \hat \lambda^2 I_{-1}(\mu){Z^{-1}_\phi} \right ]
+ f(12 \hat \lambda \Phi_c^2).
\EE
Thus $V_{1l}$ becomes finite if $\hat \lambda$ satisfies the constraint
\BE
\label{constraint}
\hat \lambda - 18 \hat \lambda^2 I_{-1}(\mu)Z^{-1}_\phi=0.
\EE
The wavefunction renormalization constant can be determined by the following
argument \cite{bran,consoli}:  The bare and renormalized two-point functions
are related by $\Gamma_B^{(2)} = Z_{\phi}^{-1} \Gamma_R^{(2)}$
and at zero momentum $\Gamma_B^{(2)}$ is given by the second derivative
of the effective potential, evaluated at the vacuum, $\varphi_c = \varphi_v$.
The renormalized inverse propagator $\Gamma_R^{(2)}$ at zero momentum is
just the renormalized Higgs mass squared, $M_H^2$, which is simply the vacuum
value of $\Omega^2$, the mass argument of the $I_1$ integral.  Hence, one has
\BE
\label{zeq}
\left. \frac {d^2 V_{1l}}{d \varphi_c^2} \right|_{\varphi_c=\varphi_{v}}
= \frac {M_H^2} {Z_{\phi}}.
\EE
On the LHS we differentiate (\ref{ol2}) twice and evaluate at $\varphi_v$
(which is where the first derivative vanishes).  On the RHS we substitute
$M_H^2 = \Omega_v^2 = 12 \lambda_B \varphi_v^2$.  This gives:
\BE
\frac{18 \lambda_B^2 \varphi^2_{v}}{\pi^2} = \frac{12 \lambda_B \varphi^2_v}
{Z_\phi}.
\EE
Together with the constraint (\ref{constraint}), this yields
\BE
Z_\phi= 12 \pi^2 I_{-1}(\mu), \qquad \hat \lambda = \frac{2}{3} \pi^2.
\EE
The renormalized potential is given by
\BE
V_{1l} =  f(12 \lambda_B \varphi_c^2) = f(8 \pi^2 \Phi_c^2),
\EE
\BE
\quad \quad \, =  \pi^2 \Phi_c^4 \left [
\ln \left( \frac{8 \pi^2 \Phi_c^2}{\mu^2} \right) - \frac{3}{2} \right ].
\EE
Note that the bare coupling constant has the form
$\lambda_B = 1/(18 I_{-1}(\mu))$
(which is the same form as in the GEP analysis \cite{aut}, except that the
coefficient here is $1/18$ instead of $1/12$).  This equation represents
``dimensional transmutation'', with $\lambda_B$ being traded for a finite
characteristic-mass-scale parameter $\mu$.  We can, of course, swap $\mu$
for the vacuum value of $\Phi_c$ (denoted $\Phi_v$).  In terms of that
parameter we have
\BE
V_{1l} = \pi^2 \Phi_c^4 \left [
\ln \left( \frac{\Phi_c^2}{\Phi^2_{v}} \right) - \frac{1}{2} \right ].
\EE

\vspace*{3mm}

{\bf 2.} This renormalization procedure can also be applied to a general CSI
theory containing scalars, fermions, and gauge fields.
In these theories all bare masses are zero, and all physical masses are
a consequence of SSB and are proportional to $\varphi_v$.
We consider the case of a massless $O(N)$-symmetric scalar sector with a
$\lambda_B (\phi^2)^2$ interaction undergoing SSB to $O(N-1)$.  The
unrenormalized 1LEP is given by \cite{Wein}:
\BE
\label{gen}
V_{1l}= \lambda_B \varphi_c^4
+ \frac{1}{2} \int_p {\rm Tr} \log(p^2 + {\cal M}_s^2)
- \frac{4}{2} \int_p {\rm Tr} \log(p^2 + {\cal M}_f^2)
+ \frac{3}{2} \int_p {\rm Tr} \log(p^2 + {\cal M}_g^2).
\EE
The coefficients 4 and 3 correspond to the number of degrees of freedom of a
Dirac fermion and a massive vector field, respectively.  The ``mass
matrices'' ${\cal M}_s$, ${\cal M}_f$ and ${\cal M}_g$ are each proportional
to $\varphi_c$.  Thus, the one-loop potential can in general be written as:
\BE
V_{1l} = \lambda_B \varphi_c^4 + \sum_i a_i I_{1}(\sqrt{ b_i \varphi_c^2}),
\EE
where $a_i$ are numerical coefficients, and $b_i$ are bare coupling constants
(or combinations of coupling constants and mixing angles).  Note that the
sum on $i$ includes the scalar contributions, so some of the $b_i$ are
proportional to $\lambda_B$.
Using (\ref{eye1s}), one can write:
\BE
\label{vv}
V_{1l} = \varphi_c^4
\left [ \lambda_B - \frac{1}{8}I_{-1}(\mu) \sum_i a_i b_i^2 \right ]
+ \sum_i a_i f(b_i \varphi_c^2).
\EE
Inserting the autonomous renormalization (\ref{autren}), supplemented by
\BE
b_i = \frac{\hat b_i}{Z_\phi},
\EE
one obtains
\BE
V_{1l} = Z_\phi \Phi_c^4 \left [\hat \lambda
- \frac{I_{-1}(\mu) \sum_i a_i \hat b_i^2}{ 8 Z_\phi} \right ]
+ \sum_i a_i f(\hat b_i \Phi_c^2).
\EE
Thus, the potential can be made finite by imposing the constraint
\BE
\label{con2}
\hat \lambda - \frac{I_{-1}(\mu) \sum_i a_i \hat b_i^2}{ 8 Z_\phi} = 0
\EE
(Cf. Eq. (\ref{constraint})).
The renormalization factor $Z_\phi$ is fixed by the relation (\ref{zeq}),
as before.  The LHS can be computed directly from (\ref{vv}), yielding
\BE
\frac{\varphi_{v}^2}{8 \pi^2} \sum_i a_i b_i^2 =
\frac{12 \lambda_B \varphi_v^2}{Z_\phi} .
\EE
Inserting (\ref{con2}) into the RHS of the last equation one obtains
$Z_\phi = 12 \pi^2 I_{-1}(\mu)$ as before. Substituting back into (\ref{con2}),
the constraint between the coupling constants becomes
\BE
\label{con3}
96 \pi^2 \hat \lambda - \sum_i a_i \hat b_i^2 = 0.
\EE
The renormalized effective potential is then given by
\BE
V_{1l}(\Phi_c) = \sum_i a_i f(\hat b_i \Phi_c^2).
\EE
Its value at its minimum is given by
\BE
V_{1l}(\Phi_v) = - \; \frac{\sum_i a_i \hat b_i^2 \Phi_v^4}{128 \pi^2}
= - \; \frac{ M^2_H \Phi_v^2}{16}.
\EE
\vspace*{3mm}

{\bf 3.} We now consider the simple example of the massless $O(N)$-symmetric
$\lambda (\phi^2)^2$ theory given by the Lagrangian
density:
\BE
\label{onphi}
{\cal L}=\frac12 \partial_\mu\phi_a\partial^\mu\phi_a
- \lambda_B (\phi_a\phi_a)^2\>,
\EE
where $a$ is summed over $1,\ldots,N$.
A non-zero classical field breaks the $O(N)$ symmetry down to
$O(N\!-\!1)$, so that there is one ``radial'' (Higgs) field and $(N\!-\!1)$
``transverse" fields.  The unrenormalized one-loop potential is
\BE
V_{1l}= \lambda_B \varphi_c^4 + I_1(\Omega) + (N-1) I_1(\omega),
\EE
where the radial and transverse masses, $\Omega$ and $\omega$ are given by
\BE
\label{om}
\Omega^2= 12 \lambda_B \varphi_c^2, \quad \quad
\omega^2= 4 \lambda_B \varphi_c^2,
\EE
so we can apply our general result with $a_1=1$, \hspace*{1mm}
$b_1 = 12 \lambda_B$, \hspace*{1mm} $a_2 = (N-1)$, \hspace*{1mm}
$b_2 = 4 \lambda_B$.  The constraint (\ref{con3}) in this case leads to
the simple equation
\BE
96 \pi^2 \hat \lambda - \hat \lambda^2(144 + 16(N-1)) = 0
\qquad \Rightarrow \qquad \hat \lambda = \frac {6 \pi^2}{N + 8},
\EE
and the renormalized effective potential is given by
\BE
V_{1l}(\Phi_c) = f(12 \hat \lambda \Phi_c^2 )
+ (N-1) f(4 \hat \lambda \Phi_c^2 ) .
\EE
\vspace*{3mm}

{\bf 4.} The same renormalization procedure can also be applied to
the $SU(2) \times U(1)$ case.  The Lagrangian density, omitting fermions, is:
\BE
{\cal L}=
-\frac{1}{4} F_a^{\mu \nu} F_{a\mu\nu}
-\frac{1}{4} B^{\mu \nu} B_{\mu\nu}
+ (D_\mu \phi)^\dagger D^\mu \phi - 4 \lambda_B (\phi^\dagger \phi)^2,
\EE
where
\BE
D_\mu = \partial_\mu + i g W_{\mu a}\frac{\tau_a}{2} +
i \frac{g'}{2} B_{\mu},
\EE
%
%Check sign of last term.
%
\BE
F^{\mu \nu}_a = \partial^\mu W_a^\nu -
\partial^\nu W_a^\mu - g \epsilon_{abc} W_b^\mu W_c^{\nu},
\EE
\BE
B^{\mu \nu} = \partial^\mu B^{\nu} - \partial^{\nu} B^{\mu}.
\EE
The Higgs field is a complex doublet, $\phi = (\phi^+, \phi^0)$, containing
four real fields, $\phi^+ = (\phi_1 + i \phi_2)/\sqrt{2}$,
$\phi^0 = (\phi_3 + i \phi_4)/\sqrt{2}$.  With this convention the
classical field is chosen to be in the $\phi_3$ direction.  Note that
$(\phi^\dagger \phi)^2 = \frac{1}{4} (\phi_a \phi_a)^2$, so the scalar
sector is just $O(4)\> \lambda (\phi^2)^2$, as in Eq. (\ref{onphi}).
The presence of a classical field induces gauge-boson mass terms in the
Lagrangian.  As usual, one must define
\BE
A^\mu \equiv \frac{g' W_3^\mu  + g B^\mu}{\sqrt{(g^2 + g'^2)}}
\equiv
\sin \theta_W W_3^\mu + \cos \theta_W B^{\mu},
\EE
\BE
Z^\mu \equiv \frac{g W_3^\mu  - g' B^\mu}{\sqrt{(g^2 + g'^2)}}
\equiv
\cos \theta_W W_3^\mu - \sin \theta_W  B^{\mu}.
\EE
in order to diagonalize these terms, which then become
\BE
\frac{\varphi_c ^2  g^2}{8} ( W_1^2 + W_2^2 ) + \frac{\varphi_c^2 (g^2 +
g'^2)}{8}Z^2.
\EE
Hence the mass parameters as a function of the classical field are given by
\BE
\Delta_W^2 = \frac{1}{4} g^2 \varphi_c^2, \qquad
\Delta_Z^2= \frac{g^2+g'^2}{4}\varphi_c^2,
\EE
with the physical masses, $M_W^2$ and $M_Z^2$, being given by these
expressions with $\varphi_c^2$ set equal to its vacuum value $\varphi_v^2$.

The one-loop potential is given by:
\BE
V_{1l}= \lambda_B \varphi_c^4 + I_1(\Omega) + 3 I_1(\omega) +
 3 \left [ 2 I_1(\Delta_W) + I_1(\Delta_Z)  \right ],
\EE
with $\Omega$ and $\omega$ as in Eq. (\ref{om}).  To renormalize the
potential, we set
\BE
g^2 = \hat g^2 / Z_\phi, \qquad g'^2 = \hat g'^2 /Z_\phi,
\EE
\BE
\lambda_B = \hat \lambda/Z_\phi,
\qquad \varphi_c = Z_\phi^{\frac{1}{2}} \Phi_c.
\EE
The constraint (\ref{con3}) in this case leads to
\BE
\label{conws}
 96 \pi^2 \hat \lambda - 192 \hat \lambda^2 - \frac{3}{16}\left( 2 \hat g^4
+ (\hat g^2 + \hat g'^2 )^2 \right) = 0 .
\EE
Since
\BE
g =\frac {e} {\sin \theta_W},  \qquad g' = \frac{e} {\cos \theta_W},
\EE
with $e^2= \hat e^2/Z_\phi$
we can simplify (\ref{conws}) and write
\BE
\label{super1}
 \pi^2\hat \lambda - 2 \hat \lambda^2 -  \frac{3 \pi^2}{96}\>
\left(\frac{\hat e^2}{4 \pi}\right)^2
\left( \frac{2 + \sec^4 \theta_W}{ \sin^4 \theta_W} \right)=0,
\EE
or, in terms of the masses:
\BE
\label{super2}
 24 \pi^2 \Phi^2_{v} M_H^2 - 4 M_H^4 - 9 (2 M_W^4 + M_Z^4) = 0.
\EE
With $\Phi_{v} = (\sqrt 2 G_F)^{-1/2} = 246$ GeV, $M_W = 80$ GeV, $M_Z=91$ GeV
\cite{rpp}, this equation leads to two possibilities for the Higgs mass,
namely:
\BE
M_H= 10 \> {\rm GeV} \qquad {\rm or } \qquad  M_H= 1.9\> {\rm TeV}.
\EE
The first possibility corresponds to a weak coupling regime where the
$\hat{\lambda}^2$ term in (\ref{super1}) is negligible.  However, a Higgs
mass below 48 GeV is now ruled out by experiment \cite{rpp}. The second
possibility corresponds to the last term in (\ref{super1}) or (\ref{super2})
being negligible, due to the smallness of the fine-structure constant
$\alpha = \hat{e}^2/(4 \pi)$.  The Higgs mass is then almost exactly what
it would be in a pure ${\rm O}(4)$ \hspace*{1mm} $\lambda (\phi^2)^2$ theory,
namely $M_H^2 = 6 \pi^2 \Phi_v^2$ $=(1.9 \; {\rm TeV})^2$.  (Note that
Ref. \cite{bran,consoli}'s result $M_H^2 = 8 \pi^2 \Phi_v^2$
$=(2.2 \; {\rm TeV})^2$ is based on single-component $\lambda \phi^4$ theory,
instead of the ${\rm O}(4)$ theory.)

The inclusion of fermions hardly changes the prediction, unless the fermions
are very heavy.  Therefore, for simplicity, we just discuss the effect of
the top quark:
\BE
{\cal L}_t = i  \bar \psi D_\mu \gamma^\mu \psi -  g_t \phi \bar{\psi} \psi.
\EE
Allowing for a colour factor of 3, the top-quark contribution to the 1LEP is
$-12 I_1(g_t \varphi_c)$.  From the general form of the constraint equation,
(\ref{con3}), and defining $g_t^2= \hat g_t^2/Z_{\phi}$, one sees that the
top quark produces an extra term $ + 12 \hat  g_t^4 $ on the LHS of Eq.
(\ref{conws}).  The new equation, in terms of the particle masses can be
written as:
\BE
 24 \pi^2 \Phi^2_{v} M_H^2 - 4 M_H^4 - 9 (2 M_W^4 + M_Z^4) + 36 M_t^4= 0 .
\EE
This equation has two real roots for $M_H^2$, as before. However, the root
corresponding to a light Higgs decreases as $M_t$ increases, and it would
give a negative $M_H^2$ if $M_t \geq 78$ GeV.
In contrast, the solution corresponding to the heavy Higgs increases slowly.
For $M_t$ below 200 GeV the effect is negligible.  For larger $M_t$'s the
Higgs mass increases more rapidly, and as $M_t \! \rightarrow \! \infty$ the
mass ratio $M_H/M_t$ tends to $\sqrt{3}$.  Thus, the Higgs is always heavier
than the heaviest fermion.

The renormalized 1LEP of the theory can be written quite compactly as:
\BE
V_{1l}(\Phi_c)=
f \left( M^2_H \frac{\Phi^2_c}{\Phi^2_v} \right)
+3 f \left(\frac{1}{3} M_H^2 \frac {\Phi_c^2} {\Phi_v^2} \right )
+6 f\left (M_W^2  \frac {\Phi_c^2} {\Phi_v^2} \right )
+3 f\left (M_Z^2  \frac {\Phi_c^2} {\Phi_v^2} \right )
- 12 f\left (M_t^2  \frac {\Phi_c^2} {\Phi_v^2} \right ).
\EE

   The 1-loop approximation can be improved by going to the Gaussian effective
potential (GEP) \cite{gepii}.  Indeed, we used to believe that the Gaussian
approximation, or some still-better nonperturbative approximation, was
necessary in order to reveal the ``autonomous'' $\lambda \phi^4$ theory
\cite{con,aut}.  However, thanks to Refs. \cite{bran,consoli}, we see that
the 1LEP does provide an adequate ``cheap substitute'' for the GEP.
We have recently calculated the GEP for the $U(1)$-Higgs model (scalar
electrodynamics) \cite{iss}, where we find remarkably good agreement between
the 1LEP and GEP results.  While we do not know the GEP for the
$SU(2) \times U(1)$ model, we can expect the Higgs mass prediction to be
governed by the O(4) $\lambda \phi^4$ sector, as was the case above.  The
GEP would give $M_H = 2.0$ TeV \cite{iss} instead of the 1LEP result, 1.9 TeV.

\vspace*{3mm}

{\bf 5.} Our analysis, being limited to the effective potential, cannot
directly answer questions about the physics at non-zero momenta.  However,
we should perhaps attempt to outline our view of the physics.   This
picture is based on the work of Consoli {\it et al} \cite{bran,consoli}, the
lattice calculations of Huang, Manousakis, and Polonyi \cite{hmp} (see also
\cite{kerman}), and our own studies of the Gaussian effective action.
The pure $\lambda \phi^4$ theory, we believe, has a {\it nontrivial} effective
potential, but has only {\it trivial} -- free particle -- excitations above
its broken-symmetry vacuum.  One can understand this as a consequence of the
infinite field re-scaling:  fluctuations that are finite on the scale of
$\varphi_c$ are infinitesimal on the scale of $\Phi_c$:  thus, they only
sample the quadratic dependence of $V_{{\rm eff}}$ in the immediate
neighbourhood of its minimum.  Three- and higher-point Green's functions are
``squelched'' by $1/Z_{\phi}^{1/2}$ factors.  In a pure O(4) $\lambda \phi^4$
theory, then, we would expect SSB, giving one, free, massive Higgs and three,
free, massless Goldstone bosons.

      In the Standard Model these Goldstone bosons are ``eaten'' and become,
essentially, the longitudinal polarization states of the $W,Z$ bosons.  The
Higgs boson is not free because it now has gauge and Yukawa couplings.
However, contrary to the conventional heavy-Higgs scenario, it would
{\it not} have strong couplings to the longitudinal $W,Z$, because the
would-be scalar self-interactions are infinitely suppressed by
$1/Z_{\phi}^{1/2}$ factors.  This means that the 2 TeV Higgs
of this picture is a relatively narrow resonance, decaying principally to
$t \bar{t}$.  It would also mean that $WW$-scattering, off resonance, is
weak.  Clearly, the question of producing and detecting such a Higgs at the
SSC needs to be studied in detail.

\vspace*{3mm}

\subsection*{Acknowledgments}
\noindent
We are most grateful to Maurizio Consoli for enlightening discussions.
\\
This work was supported by the U.S. Department of Energy under Contract
No. DE-AS05-76ER05096.

\end{document}